\newcommand{\be}{\begin{equation}}\newcommand{\ee}{\end{equation}}
\newcommand{\bea}{\begin{eqnarray}}\newcommand{\eea}{\end{eqnarray}}
\newcommand{\p}[1]{(\ref{#1})}
 \newcommand{\lb}[1]{\label{#1}}
\newcommand\s{\scriptscriptstyle}
\newcommand\di{\displaystyle}

\newcommand\cN{{\cal N}}

\newcommand\tdap{\theta^{\dal +}}

\newcommand\tdam{\theta^{\dal -}}

\newcommand{\bm}{{\bf m}}

\newcommand\ab{{\alpha\beta}}

\newcommand\diag{{\di\alpha\gamma}}

\newcommand\diar{{\di\alpha\rho}}

\newcommand\dirs{{\di\rho\sigma}}

\newcommand\pdag{\partial_\diag}
\newcommand\pdar{\partial_\diar}

\newcommand\dal{{\di\alpha}}

\newcommand\dga{{\di\gamma}}

\newcommand\drh{{\di\rho}}
\newcommand\dsi{{\di\sigma}}

\newcommand\A{{\s A}}

\newcommand\D{{\s D}}
\newcommand\R{{\s R}}

\newcommand{\2}{{\s 2}}
\newcommand{\3}{{\s 3}}
\newcommand{\4}{{\s 4}}
\newcommand{\5}{{\s 5}}

\newcommand{\pp}{{\s ++}}
\newcommand{\m}{{\s --}}

\newcommand{\Dp}{D^{\pp}}
\newcommand{\Dm}{D^{\m}}

\newcommand{\Vp}{V^\pp}
\newcommand{\Vm}{V^\m}

\documentstyle[12pt]{article}
\begin{document}
\large

\begin{center}
{\bf B.M. Zupnik}\\
\vspace{0.5cm}

{\large\bf  CONSTRAINED  SUPERPOTENTIALS IN HARMONIC GAUGE THEORIES
            WITH 8 SUPERCHARGES} \\

\end{center}
\vspace{0.5cm}

We consider $D$-dimensional supersymmetric gauge theories with 8
supercharges $(D{\leq} 6,~\cN{=}8)$  in the framework of harmonic
superspaces. The effective Abelian low-energy action for $D{=}5$ contains
the free and Chern-Simons terms. Effective $\cN{=}8$ superfield actions
for $D{\leq} 4$ can be written in terms of the superpotentials satisfying
the superfield constraints and $(6{-}D)$-dimensional Laplace equations.
The role of alternative harmonic structures is discussed.
\vspace{0.5cm}


\section{Introduction}

 The concept of  harmonic superspace $(HS)$ has been introduced
firstly for the off-shell description of matter, gauge and supergravity
superfield theories with $D{=}4,~ N_4{=}2$ supersymmetry \cite{GIK1,GI2}.
The $HS$ approach has been used also for a consistent description of
hypermultiplets and vector multiplet with $D{=}6,~ N_6{=}1$ supersymmetry
\cite{Z1}. It is convenient to use the total number of supercharges
${\cal N}{=}8$ for the classification of all these models in different
dimensions $D$ instead of the number of  spinor representations for
supercharges $N_\D$. The universality of   harmonic superspaces is
connected with the possibility of constructing $\cN{=}8$ models in
$D{ <} 6$ by a dimensional reduction. The $HS$ actions for the
hypermultiplets $q^+, \omega$ and the Yang-Mills prepotential $\Vp$ can
be described by  universal expressions in all dimensions $D \leq 6$ .
Nevertheless, ${\cal N}{=}8$ supersymmetries have some specific features
for each dimension based on differences in the structure of Lorentz groups
$L_\D$, maximum automorphism groups $R_\D$ and the set of central charges
$Z_\D$. In particular, the alternative $HS$ structures have been found for
the case $D{=}2$ \cite{IS} and $D{=}3$ \cite{Z2}. We shall study the $HS$
constructions of low-energy effective actions for the Abelian gauge
supermultiplets in $D{=}1, 2, 3$ and $5$ in terms of the analytic
prepotential $\Vp$ and the constrained superfield $(6{-}D)$-component
superfield strength $W(\Vp)$. The analogous problems have been
discussed earlier in the framework of the component-field formalism or
the formalism with the $\cN{=}4,~ D{=}1, 2, 3$ superfields (see e.g.
\cite{DE,DS,AHIS,Se}).

The main result of this work is a construction in the full $\cN{=}8$
superspace for the dimensions $D{=}1,2,3$ and 5 the  Coulomb  effective
actions  which contain the product of harmonic connections and
{\it the superpotentials}  $f_\D(W)$ satisfying the $(6{-}D)$-dimensional
Laplace equations.

Interactions of the  superfields with 8 supercharges have universal and
specific features for different dimensions, however, solutions of the most
geometric and dynamic problems can be simplified in the $HS$ formalism.
The $(6{-}D)$-dimensional Laplace equations for the low-energy
superpotentials follow from the gauge invariance in $HS$. The
non-renormalization theorems are connected with the search of
$R_\D$-invariant solutions of these equations. Analysis of the equations
for superpotentials is connected with the alternative $HS$
structures using different types of harmonics in the $D{\leq} 3,~\cN{=}8$
theories, which reflect adequately properties of duality transformations
in these models.

\section{ Effective  actions in the full and analytic superspaces}

{\bf 2.1}. Let us firstly consider the harmonic superspace with $D{=}5,~
\cN{=}8$ supersymmetry. The general five-dimensional superspace has
the coordinates $z{=}(x^\bm,~\theta^\dal_i)$, where ${\bf m}$ and $\alpha$
are the 5-vector and 4-spinor indices of the Lorentz group
$L_\5{=}SO(4,1)$, respectively, and $i$ is the 2-spinor index of the
automorphism group $R_\5{=}SU(2)$. The antisymmetric traceless $5D$
$\Gamma$-matrices $(\Gamma_\bm)_\diag$ and  invariant symplectic matrix
$\Omega_\diar$ can be constructed in terms of the  Weyl matrices and
$\varepsilon$-symbols of $SL(2,C)$.

It is convenient to consider the bispinor representation of the $5D$
coordinates and partial derivatives
\be
x^\diar={1\over2}(\Gamma_\bm )^\diar x^\bm~,\quad
\pdar={1\over2}(\Gamma^\bm )_\diar \partial_\bm~. \lb{A2}
\ee

The basic relations between the spinor derivatives $D^k_\dal$ in the
general $D=5, {\cal N}=8$ superspace have the following form:
\be
\{D^k_\dal~,~D^l_\dga\}=i\varepsilon^{kl}(\pdag
+{1\over2} \Omega_\diag Z)~, \lb{A3}
\ee
where $Z$ is the real central charge.

The $R_\5$-invariant projections of spinor derivatives $D^\pm_\dal
{=}u^\pm_iD^i_\dal$ and coordinates of the harmonic superspace
$\zeta{=}(x_\A^\bm,~\tdap),~\tdam $ can be defined with the help of the
 $SU(2)/U(1)$ harmonics $u^\pm_i$ by analogy with ref.\cite{GIK1} ($i$
is the 2-spinor index of $SU(2)$ and $q{=}\pm 1$ is the $U(1)$ charge).
The analytic Abelian prepotential $\Vp(\zeta,u)$ describes the $5D$ vector
supermultiplet, which contains the real scalar field $\phi$, the Maxwell
field $A_\bm$, the isodoublet of 4-spinors $\lambda^\dal_i$ and the
auxiliary isotriplet $X^{ik}$. The real scalar superfield of this theory
can be written in terms of the harmonic connection with the $U(1)$ charge
$-2$
\be
W= -{i\over2}D^{+\dal}D_\dal^+\int du_{\s1}
\frac{\Vp(x,u_{\s1})}{(u^+u_{\s1}^+)^2}=-2iD^{(+2)}\Vm~,\lb{A4}
\ee
where $(u^+u_{\s1}^+)^{-2}$ is the harmonic distribution \cite{GI2}.
These superfields satisfy the following constraints:
\bea
&&\Dp\Vm=\Dm\Vp~,\qquad D^{\s\pm\pm}W=0~,\lb{A5}\\
&&  D^{(+2)}_\diar W=0~,\quad\mbox{where}\;D^{(+2)}_\diar=
D^+_\dal D_\drh^+ -{1\over4}\Omega_\diar D^{+\dsi}D_\dsi^+~.
\lb{A6}
\eea

It is readily to construct the most general low-energy effective
$U(1)$-gauge action in the full $D{=}5,~\cN{=}8$ harmonic superspace
\be
S_\5=\int d^{\5}x d^{\s8}\theta du\; \Vp \Vm [ g^{-2}_\5+k_\5 W]~,\lb{A7}
\ee
where $g_\5$ is the coupling constant of dimension $1/2$, and $k_\5$ is
the dimensionless constant of the $5D$ Chern-Simons interaction.

 The low-energy $D{=}4,~\cN{=}8$ effective action conserves the
$SU(2)$ automorphism group and breaks the $U_\R(1)$ symmetry. The
corresponding $D{=}4$ superpotential $f(W,\bar{W}){=}[F(W)+\bar{F}
(\bar{W})]$ satisfies the $2D$ Laplace equation which has only holomorphic
and anti-holomorphic solutions.

{\bf 2.2}. The analogous $D{=}3,~\cN{=}8$ gauge theory can be
constructed in the superspace with the  automorphism group
$R_\3{=}SU_l(2){\times} SU_r(2)$. Coordinates of the
general superspace are $z{=}(x^\ab,~\theta^\alpha_{ia})$.
The  relations between  spinor derivatives are
\be
\{ D^{ka}_\alpha, D^{lb}_\beta\}= i\varepsilon^{kl}\varepsilon^{ab}
\partial_\ab + i\varepsilon^{kl}\varepsilon_\ab Z^{ab}~,\lb{A8}
\ee
where  $\partial_\ab{=}\partial/\partial x^\ab$ and $Z^{ab}$ are the
central charges which  commute with all generators exept for the
generators of $SU_r(2)$.

We consider
here the two-component indices $(\alpha,~\beta\ldots)$ for the
space-time group $SL(2,R)$, $(i,~k\ldots)$ for the group $SU_l(2)$ and
$(a,~b\ldots)$ for $SU_r(2)$, respectively. We shall use the
notation $u^\pm_i\equiv u_i^{\s(\pm1,0)}$ for  the harmonics of the group
$SU_l(2)$ and  $v_a^{\s(0,\pm1)}$  for the $SU_r(2)$ harmonics, and also
 the notation $D_l^{\s\pm\pm}$ for the $l$-harmonic derivatives and
 $V_l^{\s\pm\pm}$ for the $l$-version of the $3D$
harmonic gauge superfields \cite{Z2}. The notation with
two $U(1)$ charges will be introduced for the biharmonic superfields. The
gauge-covariant $SU_r(2)$-bispinor superfield of $D{=}3,~{\cal N}{=}8$
gauge theory contains the corresponding harmonic connection \cite{Z2}
\be
W^{ab}=-iD^{+\alpha a}
D^{+b}_\alpha \Vm_l~,\lb{A9}
\ee
where $D^{+b}_\alpha{=}u^+_iD^{ib}_\alpha$. This superfield
 does not depend on
harmonics in the Abelian case.

The $SL(2,R){\times} SU_l(2)$ invariant Coulomb effective action can be
expressed in terms of the  superpotential $f_\3(W^{ab})$
\be
 S_\3=\int d^{\s3}x d^{\s8}\theta du\; \Vp_l \Vm_l f_\3(W^{ab})~.\lb{A12}
\ee
The gauge invariance produces the following constraint:
\be
D^{+c}_\alpha D^+_{c\beta}\, f_\3(W^{ab})=0.\lb{A13}
\ee

This constraint is equivalent to the three-dimensional Laplace equation
\be
\Delta_\3^w f_\3(W^{ab})=0~,\qquad
\Delta_\3^w=\frac{\partial}{\partial W^{ab}}
\frac{\partial}{\partial W_{ab}}
~.\lb{Lapl3}
\ee

The $R_\3$-invariant solution of Eq.\p{Lapl3} has the following form:
\be
f^\R_\3(w_\3)= g^{-2}_\3+ k_\3 w_\3^{-1}~,\qquad w_\3=
\sqrt{W^{ab}W_{ab}}
\lb{A14}
\ee
where $g_\3$ is the coupling constant of dimension $d{=}{-}1/2$, and
$k_\3$ is the dimensionless constant of the ${\cal N}{=}8$ $WZNW$-type
interaction of the vector multiplet. This interaction is well defined for
nonzero values of the central-charge modulus $|Z|_\3{=}\sqrt{Z^{ab}
Z_{ab}}$, when its decomposition in terms of $\hat{W}_{ab}{=}W_{ab}-
Z_{ab}$ is nonsingular. It should be remarked that the  superfield
interactions of the $3D$-vector multiplets with dimensionless constants
(Chern-Simons terms) have been constructed earlier for the case
${\cal N}{=}4$ \cite{ZP} and ${\cal N}{=}6$ \cite{ZK,Z2}.

The  constraints  on the superfield $W^{ab}$ can be interpreted as an
alternative $r$-analyticity  of the following projection of this
superfield in  the biharmonic  superspace $(BHS)$:
\be
W^{\s(0,2)}=v_a^{\s(0,1)}v_b^{\s(0,1)} W^{ab}=-iD^{\s(2,2)}V_l^{\s(-2,0)}
(x,u)=-i\int du D^{\s(-2,2)}V_l^{\s(2,0)}(x,u)~.\lb{A15b}
\ee
where $V_l^{\s(\pm2,0)}{\equiv}V_l^{\s\pm\pm}$ and $D^{\s(\pm2,2)}{=}
 u_i^{\s(\pm1,0)}u_k^{\s(\pm1,0)}v_a^{\s(0,1)}v_b^{\s(0,1)}
D^{ia\alpha} D^{kb}_\alpha$.

The  $3D$-superpotential \p{Lapl3} can be written in the form of the
integral over the harmonics $v_a^{\s(0,\pm1)}$:
\be
f_\3(W^{ab}) =\int dv F_\3[W^{\s(0,2)},v^{\s(0,\pm1)}]~,\lb{altharm3}
\ee
where $F_\3$ is an arbitrary function with $q{=}(0,0)$. The effective
action in the full superspace \p{A12} can be transformed to
the equivalent representation in  the $r$-analytic superspace
\be
S_\3=\int d^\3x D^{\s(0,-4)} dv [W^{\s(0,2)}]^2 F_\3[W^{\s(0,2)},~
v^{\s(0,\pm1)}]~,\qquad D^{\s(0,-4)}=D^{\s(2,-2)}
D^{\s(-2,-2)}~.\lb{A15c}
\ee

The mirror symmetry connects the $l$-vector multiplet $W^{\s(0,2)}
(V^{\s(2,0)})$ with the $r$-analytic hypermultiplet $\omega_r$.

{\bf 2.3}. The two-dimensional (4,4) superfields and
the corresponding $\sigma$-models have been discussed in the
ordinary superspace \cite{GHR,GoI} and in the framework of  alternative
harmonic superspaces \cite{IS}. The (4,4) gauge theory has been considered
in the formalism of the (2,2) superspace \cite{DS}. We shall study the
geometry of this theory in the manifestly covariant harmonic formalism. In
the (4,4) superspace with the coordinates $(y, \bar{y}, \theta^{i\alpha},
\bar{\theta}^{ia})$, we shall use the automorphism group $R_\2{=}SU_c(2)
{\times}SU_l(2){\times}SU_r(2)$ ( the  notation of spinor indices for
these  groups: $ c)~ i, k, \ldots$; $l)~\alpha,\beta\ldots$ and $r)~a,
b\ldots$, respectively).
The algebra of spinor derivatives in this superspace is
\bea
&& \{D_{k\alpha},D_{l\beta}\}=\varepsilon_{kl}\varepsilon_\ab\partial_y~,
\lb{A16b}\\
&& \{\bar{D}_{ka},\bar{D}_{lb}\}=\varepsilon_{kl}\varepsilon_{ab}
\bar{\partial}_y ~,\lb{A16c}\\
&& \{D_{k\alpha},\bar{D}_{lb}\}=i\varepsilon_{kl}Z_{\alpha b}~,\lb{A16d}
\eea
where $Z_{\alpha b}$ are the central charges.

The superfield constraints of the nonabelian (4,4) gauge theory are
\bea
&& \{\nabla_{k\alpha},\nabla_{l\beta}\}=\varepsilon_{kl}\varepsilon_\ab
   \nabla_y~,\lb{A16}\\
&& \{\bar{\nabla}_{ka},\bar{\nabla}_{lb}\}=\varepsilon_{kl}
\varepsilon_{ab} \bar{\nabla}_y~,\lb{A17}\\
&&\{\nabla_{k\alpha},\bar{\nabla}_{lb}\}= i\varepsilon_{kl} W_{\alpha b}~,
\lb{A18}
\eea

The authors of ref.\cite{IS} have discussed the three types  of harmonics:
$u^\pm_i=u^{\s(\pm1,0,0)}$ for $SU_c(2)/U_c(1)$ ; $l_\alpha^{\s(0,\pm1,0)
}$ for $SU_l(2)/U_l(1)$; and $r_a^{\s(0,0,\pm1)}$ for $SU_r(2)/U_r(1)$
(in our notation). The basic geometric structures of the gauge theory are
connected mainly with the harmonics $u^\pm_i$  and the corresponding
analytic coordinates $\zeta_c=(y_c,~\theta^{+\alpha})$ and $\bar{\zeta}_c
=(\bar{y}_c,~\bar{\theta}^{+a})$. The $SO(4)$-vector superfield strength
$w_m$ for the $2D$ analytic gauge prepotential $\Vp_c(\zeta_c~,
\bar{\zeta}_c,~u)$ can be constructed by analogy with $D{=}3$
\be
W_{\alpha b}\equiv (\sigma^m)_{\alpha b}W_m=-iD^+_\alpha \bar{D}^+_b
\Vm_c \lb{A19}
\ee
where $(\sigma^m)_{\alpha b}$ are the $SO(4)$ Weyl matrices.
$W_{\alpha b}$ satisfies the superfield constraints analogous to the
constraints of the so-called twisted  multiplet \cite{GHR}.

In the full  (4,4) superspace, one can construct the  effective
 action of the $U(1)$ gauge theory
\be
S_\2=\int d^{\2}x d^{\s8}\theta du\; \Vp_c \Vm_c f_\2(W_m)~,\qquad
(D^+)^2 f_\2(W_m)=(\bar{D}^+)^2 f_\2(W_m)=0~.
\lb{A22}
\ee

The general (4,4) superpotential satisfies the $4D$ Laplace equation
\be
\Delta_\4^w f_\2(W_m)=0~,\qquad
\Delta_\4^w=\frac{\partial}{\partial W_m}
\frac{\partial}{\partial W_m}
~.\lb{Lapl4}
\ee

The $R_\2$-invariant solution for the (4,4) superpotential is determined
uniquely
\be
f^\R_\2(w_\2)= g^{-2}_\2+ k_\2 w_\2^{-2}~,\quad w_\2=
\sqrt{W_mW_m}~.\lb{A23}
\ee
The analogous function has been considered in the derivation of the
$R_\2$-invariant (2,2) K\"{a}hler potential of the $D{=}2, (4,4)$ gauge
theory \cite{DS}. The manifestly (4,4) covariant formalism of the
harmonic gauge theory simplifies the proof of the non-renormalization
theorem.

The biharmonic  representation of the (4,4) superpotential is
natural for the solutions of Eq.\p{Lapl4}
\be
f_\2(W^{\alpha a})=\int dl dr F_\2[W^{\s(0,1,1)},
~l,~r]~,\qquad W^{\s(0,1,1)}=l_\alpha^{\s(0,1,0)}
r_a^{\s(0,0,1)}W^{\alpha a}~, \lb{trhar}
\ee
where $F_\2$ is a real $rl$-analytic function with the zero
$U(1)$-charges.

This  projection of the vector multiplet \p{A19} satisfies the conditions
of the $rl$-analyticity in the triharmonic superspace
\bea
&&u_i^{\s(\pm1,0,0)}l_\alpha^{\s(0,1,0)}D^{i\alpha} W^{\s(0,1,1)}\equiv
D^{\s(\pm1,1,0)}
W^{\s(0,1,1)}=0~,\lb{A24a}\\
&&u_i^{\s(\pm1,0,0)}r_a^{\s(0,0,1)}\bar{D}^{ia}W^{\s(0,1,1)}\equiv
 \bar{D}^{\s(\pm1,0,1)} W^{\s(0,1,1)}=0 \lb{A24}
\eea
and the harmonic conditions
\be
D_c^{\s(\pm2,0,0)} W^{\s(0,1,1)}=D_l^{\s(0,2,0)}W^{\s(0,1,1)}=
D_r^{\s(0,0,2)}W^{\s(0,1,1)}=0 \lb{A24b}
\ee
which are analogous to the constraints on the $q^{\s(1,1)}$ superfield
of ref.\cite{IS} (this notation does not indicate the $U_c(1)$ charge).
 Note that the vector multiplet $W^{\s(0,1,1)}$ contains the
$2D$ vector field instead of the auxiliary scalar component in
the superfield $q^{\s(1,1)}$.

Using Eqs.\p{trhar} and (\ref{A22}) one can obtain the following
equivalent representation of the effective (4,4) action in the
$rl$-analytic superspace:
\be
S_\2=\int dl dr d^{\2}x D^{\s(1,-1,0)}D^{\s(-1,-1,0)}\bar{D}^{\s(1,0,-1)}
\bar{D}^{\s(-1,0,-1)}[W^{\s(0,1,1)}]^2 F_\2[W^{\s(0,1,1)},~l,~r]~,
\lb{A24c}
\ee
where
\be
W^{\s(0,1,1)}=-iD^{\s(1,1,0)}\bar{D}^{\s(1,0,1)}V^{\s(-2,0,0)}=-i
\int du D^{\s(-1,1,0)}\bar{D}^{\s(-1,0,1)}V^{\s(2,0,0)}~.\lb{2dvm}
\ee

The  action of the $q^{\s(1,1)}$ multiplet with an analogous
structure of the (4,4) $\sigma$-model  has been constructed
in ref.\cite{IS}. This multiplet is dual to the $rl$-analytic
multiplet $\omega^{\s(\pm1,\mp1)}$.

{\bf 2.4}. The one-dimensional $\sigma$-models have been considered in the
$\cN{=}4$ superspace \cite{BP,ISm}. Recently, this superspace has
been used also for the proof of the non-renormalization theorem in the
$\cN{=}8$ gauge theory \cite{DE}.

We shall consider the $D{=}1,~\cN{=}8$ superspace which is based on
the automorphism group $R_{\s1}{=}SU_c(2){\times} Spin(5)$ and has the
coordinates $(t,~\theta^\dal_i) $, where $i$ is the 2-spinor index and
$\alpha$ is the 4-spinor index of the group $Spin(5){=}USp(4)$.
 The algebra of spinor derivatives is
\be
\{D^k_\dal,D^l_\drh\}=i\varepsilon^{kl}\Omega_\diar\partial_t
+i\varepsilon^{kl}Z_\diar~,\lb{A26}
\ee
where $Z_\diar$ are central charges.

 Constraints of the $1D$ vector multiplet correspond to the integrability
conditions of the $c$-analyticity.
The
analytic $1D$ coordinates $\zeta_c{=}(t_c,~\theta^{+\dal})$ can be
defined via the standard harmonics $u^\pm_i{\equiv}u^{\s(\pm1,0,0)}_i$.
The notation and algebra of the harmonized spinor derivatives
$ D^\pm_\dal$ are similar for $D{=}1$ and $D{=}5$ cases. The superfield
strength for the corresponding harmonic gauge connections $V_c^{\s\pm\pm}$
is the 5-vector $W_\bm$ (or traceless bispinor $W_\diar$) with respect to
$Spin(5)$
\be
W_\diar\equiv {1\over2}(\Gamma^\bm)_\diar W_\bm=-iD^{(+2)}_\diar\,\Vm_c~,
\quad D^{\s(+2)}W_\diar =0~,
\lb{A25}
\ee
where the $Spin(5)\quad\Gamma$-matrices and  notation \p{A6} are used.

In the Abelian gauge group, four components of this bispinor are twisted
superfields, for example, $D^\pm_1 W_{13}{=}D^\pm_3 W_{13}{=}0$.

The effective action has the following form in the full $D=1$ superspace:
\be
S_{\s1}=\int dt d^{\s8}\theta du\; \Vp_c \Vm_c\; f_{\s1}(W_\bm)~.
\lb{A27}
\ee

The gauge invariance of $S_{\s1}$ is equivalent to the $5D$ Laplace
equation for the superpotential
\be
D^{(+2)}\,f_{\s1}(W_\bm) =0~\rightarrow~ \Delta_\5^w f_{\s1}(W_\bm) =0~.
 \lb{5Dcon}
\ee

The $R_{\s1}$-invariant $D{=}1$ superpotential
\be
f^\R_{\s1}(w_{\s1})= g^{-2}_{\s1}+ k_{\s1} w_{\s1}^{-3}\lb{A28b}
\ee
is determined via the length of the 5-vector
\be
 w_{\s1}=( W^\dirs W_\dirs)^{1/2}~. \lb{A28}
\ee

The biharmonic construction of the general $5D$ superpotential \p{5Dcon}
requires the use of  harmonics $v_\dal^{(0,\pm1,0)},~v_\dal^{(0,0,\pm1)}$
of the group $USp(4)$ \cite{IKNO} and the coorresponding
harmonic projection of the bispinor superfield
\bea
&& f_{\s1}(W^\diar)=\int dv F_{\s1}[W^{(0,1,1)},v_\dal]~,\lb{A29}\\
&&W^{(0,1,1)}=v_\dal^{(0,1,0)}v_\drh^{(0,0,1)}W^\diar~,
\lb{A30}
\eea
where the real function $F_{\s1}$ of the superfield
$W^{(0,1,1)}$ and $v$-harmonics is considered.
The constraints of the superfield $W^\diar$ are equivalent to the
conditions of the $v$-analyticity
\be
u^{\s(\pm1,0,0)}_iv_\dal^{(0,1,0)}D^{i\dal}W^{(0,1,1)}=
u^{\s(\pm1,0,0)}_iv_\dal^{(0,0,1)}D^{i\dal}W^{(0,1,1)}=0\lb{A31}
\ee
together with the harmonic conditions in $u$- and $v$-variables
\bea
&&D_c^{\s(\pm2,0,0)}W^{(0,1,1)}=D_v^{\s(0,2,0)}W^{(0,1,1)}=0~,\lb{A32}\\
&&D_v^{\s(0,0,2)}W^{(0,1,1)}=D_v^{\s(0,1,1)}W^{(0,1,1)}=0~.\lb{A33}
\eea

By analogy with $D{=}2$ and 3 we can analyze the
equivalent form of $S_{\s1}$ in the $v$-analytic superspace and
corresponding duality transformations.

I am grateful to E.A. Ivanov for  stimulating discussions.
This work is partially supported  by  grants  RFBR-96-02-17634,
RFBR-DFG-96-02-00180,  INTAS-93-127-ext and INTAS-96-0308, and
by  grant of the Uzbek Foundation of Basic Research N 11/97.

\end{document}